\documentstyle[preprint,aps, psfig]{revtex}
\newcommand{\beq}{\begin{equation}}
\newcommand{\eeq}{\end{equation}}

\begin{document}

\title{Comments on P$^3$M, FMM, and the Ewald Method for Large Periodic
       Coulombic Systems}

\author{E. L. Pollock and Jim Glosli}
\address{Physics Department, Lawrence Livermore National Laboratory,
         University of California, Livermore, California 94550 }

\date{\today}
\maketitle

\begin{abstract}

    Prompted by the need to simulate large molecular or gravitational
 systems and the
availability of multiprocessor computers, alternatives to the standard
Ewald calculation of Coulombic interactions have been developed.
The two most popular alternatives, the fast multipole method (FMM) and
the
particle-particle particle-mesh (P$^3$M) method are compared here to
the Ewald method for a single processor machine. Parallel
processor implementations of the P$^3$M and Ewald methods are
compared. The P$^3$M method is found to be both faster than the
FMM and  easier to implement
efficiently as it relies  on commonly available software (FFT
subroutines).
Both the Ewald and  P$^3$M  method are easily implemented on parallel
architectures with the P$^3$M method the clear choice for large
systems.
\end{abstract}

\section{Introduction}

   The evaluation of Coulombic interactions for large systems is a
common
computational problem. In biomolecular systems the scale of the
structures,
for example biological membranes, often require simulating large
systems.
Many algorithms have been used for this. Here three of the most common,
the Ewald \cite{ew1}, particle-particle particle-mesh, P$^{3}M$
\cite{hockney},
 and fast multipole method, FMM, \cite{fmm1} are commented on and
compared
for single processor
and (for the first two methods) multiprocessor computers. Our aim
is to suggest areas of application for each method and provide some
guides to
their implementation.

   Our observations are the result of applying these methods to
condensed matter problems rather than a mathematical or algorithmic
interest in the methods themselves. This explains some of the omissions
and qualitative nature of much of the discussion.
For example the implementation for multiprocessor computers was
directed
to the study of molecular systems where the immediate goal was to
evaluate the
long range interactions in a time comparable to the short range
interactions.

  Section two presents the Ewald and  P$^3$M \cite{luty}, \cite{henrik}
methods
together since they are very similar. Appendices A and B give a
compilation of necessary formulas for the P$^3$M method and a
discussion of parameter selection.

 Section three, together with appendix C, discusses the FMM method
and how it can be efficiently implemented on single processor machines.
Although
the operations count for this method scales linearly with the number of
charges, different codes have shown a variation of two orders of
magnitude
in the number of charges required for this method to  exceed Ewald in
speed.

      Section four discusses single processor implementations of the
three
methods starting with a discussion of accuracy for P$^3$M methods.
Timings for the three methods are then compared. Finally the relative
advantages of these methods in treating  non-cubic
periodic cells, alternate boundary conditions, non-Coulombic
interactions, and two dimensional systems are considered.

 Section five presents parallel implementations of the Ewald and
P$^3$M methods (specifically on the CRAY T3D).
\section{ The P$^3$M and the Ewald Method}

        The P$^3$M method is closely related to the Ewald method so we
consider the two together here to give the usual heuristic derivation
of both
methods.

        The total electrostatic potential energy for a system of $N$
 point charges
\beq
U={1\over 2}\sum_{i=1}^{N}\sum_{j\neq i}^{N}
  \begin{array}{c} \underline{ Z_{i}Z_{j}}\\r_{ij}   \end{array}
                   \label{eq1.1}
\eeq
is rewritten by
adding and subtracting  a term corresponding physically to the
electrostatic
energy of a system of  smooth spherical
charges, with a density $\hat{\rho}_{i}(r)$,
centered on the particle positions to obtain \cite{ewcomment}

\beq
\begin{array}{lll}

U&=&\displaystyle {1\over 2}\sum_{i=1}^{N}\sum_{j\neq i}^{N}\left[
\begin{array}{c} \underline{ Z_{i}Z_{j}}\\r_{ij}  \end{array}
-{\bf \int}\int\begin{array}{c}\underline{\hat{\rho}_{i}(r)
\hat{\rho}_{j}(r')}
    \\|r-r'|\end{array} drdr'\right]\\
&&\displaystyle
+{1\over
2}\sum_{i=1}^{N}\sum_{j=1}^{N}\int\int\begin{array}{c}\underline{
\hat{\rho}_{i}(r) \hat{\rho}_{j}(r')}\\|r-r'|\end{array}
-{1\over
2}\sum_{i=1}^{N}\int\int\begin{array}{c}\underline{\hat{\rho}_{i}(r)
\hat{\rho}_{i}(r')}\\|r-r'|\end{array}

\end{array}
                   \label{eq1.2}
\eeq

   The first, bracketed, term now corresponds to particles interacting
through
a short-ranged interaction which is zero beyond the overlap of
 $\hat{\rho}_{i}$ $\hat{\rho}_{j}$.
The second term corresponds to the Coulomb energy of a smooth charge
distribution $\hat{\rho}(r)\equiv \sum_{i}^{N}\hat{\rho}_{i}(r)$ and
the
last term is a constant self-energy.

   The Ewald formula uses  a Gaussian for $\hat{\rho}_{i}(r)$
\beq
\hat{\rho}_{i}(r)=Z_{i}( G^{2}/\pi )^{3/2}
  \exp [-G^{2}(r-r_{i})^2]\;.
                   \label{eq1.3}
\eeq
The P$^3$M method allows any choice for $\hat{\rho}_{i}$ but we have
found
no advantage in the usual alternative choices and will use the Gaussian
form
throughout. The interaction of Gaussian shaped charges  can be
evaluated
analytically and gives rise to an error
function. Doing this the potential energy becomes
\beq
U={1\over 2}\sum_{i=1}^{N}\sum_{j\neq i}^{N}
\begin{array}{c} \underline{ Z_{i}Z_{j}}\\r_{ij}  \end{array}
erfc\left( G r_{ij}/\sqrt{2}\right)
+{1\over 2}\int\int\begin{array}{c}\underline{
\hat{\rho}(r) \hat{\rho}(r')}\\|r-r'|\end{array}drdr'
-{G\over \sqrt{2\pi}}\sum_{i=1}^{N}Z_{i}^{2}
                   \label{eq1.4}
\eeq

   The treatment of the remaining term now distinguishes P$^3$M from
Ewald.
The Ewald formula results from an exact, analytic evaluation of the
term
\beq
\int\int\begin{array}{c}\underline{
\hat{\rho}(r) \hat{\rho}(r')}\\|r-r'|\end{array}drdr'=
\Omega\sum_{k\neq 0} {4\pi\over k^{2}}\hat{\rho}(k)\hat{\rho}(-k)
                   \label{eq1.5}
\eeq
which together with
\beq
\hat{\rho}(k)=\sum_{i=1}^{N}{1\over \Omega}\int e^{ik\cdot r}
  Z_{i}\left( G^{2}/\pi\right)^{3/2}e^{-G^{2}(r-r_{i})^2}dr=
{1\over\Omega}\exp(-k^{2}/4G^{2}) S(k)\;,
                   \label{eq1.6}
\eeq
where the charge structure factor
$S(k)\equiv\sum_{i=1}^{N}Z_{i}e^{ik\cdot r_{i}}$, gives
\beq
U={1\over 2}\sum_{i=1}^{N}\sum_{j\neq i}^{N}
\begin{array}{c} \underline{ Z_{i}Z_{j}}\\r_{ij}  \end{array}
erfc\left(
G r_{ij}/\sqrt{2}\right)-{G\over \sqrt{2\pi}}\sum_{i=1}^{N}Z_{i}^{2}
+{1\over 2}\sum_{k\neq 0}
{4\pi\over \Omega }
\begin{array}{c}\underline{\exp(-k^{2}/2G^{2})}\\k^{2}\end{array}\;
                  |S(k)|^{2}\;.
                   \label{eq1.7}
\eeq
$\Omega$ is the periodic system volume.

      A prescribed accuracy requires a cutoff $r_{c}\propto 1/G$,
so $O(N\rho r^{3}_{c})\sim O(N^{2}/G^{3}\Omega)$
operations for the first term in the above equation, and
$k_{max}\propto G$
for the last term so $O(G^{3}\Omega)$ $k$ vectors and $O(NG^{3}\Omega)$
 operations, since computing each
$S(k)$ requires $O(N)$ operations.
Varying $G^{3}\Omega$ to minimize the total number of operations gives
the
optimal $G^{3}\Omega\propto\sqrt{N}$ and the familiar
$N^{3/2}$ scaling of computation time with the number of charges.

     The P$^3$M  method follows from treating   eqn. \ref{eq1.5}
by numerical methods. Essentially the density $\hat{\rho}(r)$ is
assigned
to a grid and then $\hat{\rho}(k)$ computed by an FFT.
Computing the electrostatic forces on each particle
\beq
{\bf F}_{i}=-{\bf \nabla}_{i}U=-\int
\hat{\rho}_{i}(r){\bf\nabla}\hat{\Phi}(r) dr
                 \label{eq1.8}
\eeq
also requires transforming the field $i{\bf k}\hat{\Phi}(k)$ back to
 real space.

     Again evaluating the first (``real space'') term in eqn.
\ref{eq1.4}
to a given precision
requires $r_{c}\propto 1/G$ or $O(N^{2}/G^{3}\Omega)$ operations.
For an accurate numerical evaluation of eqn. \ref{eq1.5} (the
``reciprocal
space'' term) a grid that accurately resolves $\hat{\rho}_{i}$ is
necessary.
{}From eqn.~\ref{eq1.3} the width of $\hat{\rho}_{i}\approx1/G$ so a
grid spacing $\Delta$ such that
$1/G\Delta\approx n$, where $n$ is
some number (say 8 or greater), is needed.
 This gives $G^{3}\Omega n^{3}$ grid points.
The FFT to solve the Poisson equation uses $O(G^{3}\Omega n^{3}
\ln(G^{3}\Omega n^{3}))$ steps and forming the density
takes  $O(N)n^{3}$ steps.
Varying $G^{3}\Omega$ to minimize the time gives the optimal
$G^{3}\Omega\propto N$ and time scaling of $O(Nln(N))$.
If other considerations govern the choice of $G$ then P$^3$M still
scales as
$O(Nln(N))$ while Ewald deteriorates to $O(N^{2})$.
Note also that the optimal $G$ now is constant as $N$ increases at
fixed
density. By contrast the optimal $G$ for the Ewald method decreases at
constant density as $N^{-1/6}$ implying a longer ranged ``real space''
interaction and thus memory for neighbor tables increasing as
$N^{3/2}$.

    This straight forward implementation of P$^3$M, referred to below
as
primitive P$^{3}$M,  is practical and
clearly far preferable to Ewald evaluations for large systems.
It is found (as already suggested by the choice $n\approx 8$ above)
 however that the density of a single particle, $\hat{\rho}_{i}(r)$,
must often be spread over several hundred grid points. Clearly
distributing
each charge to fewer grid points would yield a still faster algorithm.

 Hockney and Eastwood \cite{hockney} suggested using a
different, narrower, density (or assignment function) for this last
term and
compensating by modifying the Coulomb Green's function.
The basic idea may be seen in rewriting eqn.~\ref{eq1.5}
\beq
\Omega\sum_{k\neq 0} {4\pi\over k^{2}}|\hat{\rho}(k)|^{2}
=\Omega\sum_{k\neq 0} {4\pi\over k^{2}}
\begin{array}{c}\underline{|\hat{\rho}(k)|^{2}}\\ |W(k)|^{2}\end{array}
|W(k)|^{2}
                   \label{eq1.9}
\eeq
where the ``assignment function''  $W(r)=\sum_{i=1}^{N}W_{i}(r)$
would be narrower than $\hat{\rho}_{i}$ making it easier to form the
``density'' and to interpolate the forces. The Coulomb Green's function
is thus
modified to
\beq
 {4\pi\over k^{2}}
\longrightarrow {4\pi\over k^{2}}
\begin{array}{c}\underline{|\hat{\rho}(k)|^{2}}
 \\ |W(k)|^{2}\end{array}\;.
                    \label{eq1.10}
\eeq

  Since this exact compensation requires  $\hat{\rho}(k)$ and $W(k)$
which vary
for each configuration nothing has been gained.
Instead Hockney
and  Eastwood use a modified Coulomb Green's function which minimizes
the mean
squared error in the forces (due to the new assignment function and
also
finite size grid errors) for charges uniformly distributed in the cell.
The resulting formula as well as those for the assignment functions are
given in Appendix A. Details may be found in Hockney and Eastwood or
more concisely in \cite{ferrell} and we give results below
demonstrating
the final errors associated with various assignment schemes.

      The steps in the $P^{3}M$ method can now be summarized as:
\begin{itemize}
\item Compute the short ranged terms.
\item Form an effective density $W(r)=\sum_{i=1}^{N}W_{i}(r)$, where
      specific forms for various $W_{i}(r)$ which assign the density to
      $n=3,4,5,\ldots$ grid points in each dimension
      are given in appendix A.
\item Using the modified Coulomb Green's function, also given in
       appendix A, solve Poisson's equation to get the potential and
      electric fields due to the effective density.
\item Finally, interpolate the fields back to the particles ( eqn.
\ref{eq1.8})
      using the $W_{i}(r)$.
\end{itemize}

      A discussion on the selection of the assignment order, grid size,
and
parameter $G$ is given in appendix B.
\section{Description of FMM}

      The Fast Multipole Method primarily due to Greengard and
Rokhlin~\cite{fmm1} has been
described many times. A concise description is given in a paper
discussing its
first three dimensional implementation \cite{schmidt} and  an intuitive
overview of the reasoning behind the steps in ref.~\cite{greengard2}.
Here we review this description and stress the efforts needed to make
it
efficient. The importance of the method is its $O(N)$ scaling with the
number of charges.

      The FMM method for calculating Coulomb interactions is based on
two related expansions: the multipole expansion
\beq
 V({\bf r})=4\pi\sum_{l,m}^{lmax}
    \begin{array}{c}\underline{M_{lm}}\\ (2l+1) \end{array}
    \begin{array}{c}\underline{Y_{lm}(\hat{\bf r})}\\
r^{l+1}\end{array}
     +O(r_{i}^{max}/r)^{lmax+1}\;,
             \label{eq3.1}
\eeq
      where the multipole moment for $N$ charges is
\beq
M_{lm}=\sum_{i}^{N}q_{i}r_{i}^{l}Y_{lm}^{*}(\Omega_{i})\;,
             \label{eq3.2}
\eeq
     which converges for ${\bf r}>\max{\{\bf r}_{i}\}$,
and the local expansion
\beq
 V({\bf r})=4\pi\sum_{l,m}^{lmax} L_{lm} r^{l}Y_{lm}(\hat{\bf r})
            +O(r/r_{i}^{min})^{lmax+1}
             \label{eq3.3}
\eeq
    where
\beq
L_{lm}=\sum_{i}{q_{i}\over (2l+1)}
           {Y^{*}_{lm}(\Omega_{i})\over r_{i}^{l+1}}
             \label{eq3.4}
\eeq
which converges for ${\bf r}<\min{\{\bf r}_{i}\}$.
The purpose of the FMM is to calculate the local expansion coefficients
due to charges at some distance from the point of interest and to
account
for the closer charges by a direct summation.

     This is done in five steps. The simulation cell is successively
subdivided. The largest division is the cell itself. This is divided
into say eight cubes (for example). Each of these cubes is then
subdivided and so on for a prescribed number of subdivisions, $L$,
so that at the finest subdivision there are $8^{L}$ cells.
The five steps are now:
\begin{enumerate}
\item Compute multipole moments (using eqn.~\ref{eq3.2}) for all
      cells at final level of subdivision (smallest cells);
\item  Sweep up from smallest cells to largest cell to get multipole
       moments for cells at all subdivision levels using the formula
       to shift the origin of a multipole expansion given below;
\item Sweep down from largest (single) cell to smallest cells to get
      local expansion coefficients in the smallest boxes according to
      the repeated sequence;
\begin{description}
\item[a] Transform local expansion of larger cell to cells at next
level
         of subdivision using the formula for shifting the origin of a
         local expansion given below;
\item[b]  Add to these local expansion coefficients the contribution
         from cells at next level of subdivision which have not been
          included yet and which are not near neighbors of the cell
         being considered. This uses the multipole moments of these
cells
         according to a formula given below. The first time this is
done
         non-nearest
         neighbor images of the simulation cell will be included for
the
         case of periodic boundary conditions.
\end{description}
\item Once the preceding step has reached the finest subdivision level
      evaluate the potential and fields for each particle using the
      local expansion coefficients for the (smallest) cell containing
      the particle.
\item  Add the contributions from other charges in the
      same cell  and in near
      neighbor cells (their contribution is not in the local expansion
      coefficients) by direct summation.
\end{enumerate}

       The algebra for these steps consists of convolution type
sums. With the notation
\beq
 a_{lm}=\begin{array}{c}\underline{(-1)^{l+m}\sqrt{2l+1}}\\
       \sqrt{4\pi (l+m)! (l-m)!}\end{array}
                 \label{eq3.5}
\eeq
 and replacing the multipole moments and local expansion coefficients
by
\beq
{\cal M}_{lm}\equiv {a_{lm}M_{lm}\over (2l+1)}
                 \label{eq3.6}
\eeq
\beq
{\cal L}_{lm}\equiv {(2l+1)L_{lm}\over a_{lm}}
                 \label{eq3.7}
\eeq
the needed formulae for the steps are \cite{remark}:
\begin{enumerate}
\item Calculate smallest cell multipole moments from the definition;
\item for shifting the origin of a multipole expansion
\beq
     {\cal M}^{'}_{l'm'}=\sum_{l=0}^{l'}\sum_{m=-l}^{l}
                          t^{MM}(l'-l,m'-m){\cal M}_{lm}
                 \label{eq3.8}
\eeq
\beq
     t^{MM}(l,m)=\begin{array}{c}\underline{4\pi (-1)^{l}a_{lm}}\\ 2l+1
                 \end{array} Y^{*}_{lm}(\hat{\bf R}) R^{l}
                 \label{eq3.9}
\eeq
     where ${\bf R}$ points from the old to new cell origin;
\item for shifting the origin of a local expansion use
\begin{description}
\item[a]
\beq
     {\cal L}^{'}_{l'm'}=\sum_{l=l'}^{l_{max}}\sum_{m=-l}^{l}
                           t^{LL}(l-l',m-m'){\cal L}_{lm}
                 \label{eq3.10}
\eeq
\beq
     t^{LL}(l,m)=\begin{array}{c}\underline{4\pi a_{lm}}\\ 2l+1
                 \end{array} Y_{lm}(\hat{\bf R}) R^{l}
                 \label{eq3.11}
\eeq
      and for adding multipoles to a local expansion use
\item[b]
\beq
      {\cal
L}^{'}_{l'm'}=(-1)^{l'+m'}\sum_{l=0}^{l_{max}}\sum_{m=-l}^{l}
                           t^{LM}(l+l',m'-m){\cal M}_{lm}
                 \label{eq3.12}
\eeq
\beq
     t^{LM}(l,m)=\begin{array}{c}\underline{4\pi (-1)^{m}}\\ a_{lm}
                 \end{array} Y^{*}_{lm}(\hat{\bf R})/ R^{l+1}
                 \label{eq3.13}
\eeq
     where again ${\bf R}$ points from the origin of the multipole
     calculation to the origin for the local expansion.
\end{description}
\end{enumerate}
     The spherical harmonics are those of Jackson's
{\em Classical Electrodynamics} \cite{jackson}.
      A discussion of the efficient implementation of this algorithm is
given
in appendix C.

\section{Numerical results and comparison of the methods on a single
processor}
\subsection{Comments on accuracy of the P$^{3}$M  method
(discretization error)}

      Although an extended discussion of P$^{3}$M accuracy is found in
ref. \cite{hockney} we present here some indicative results to aid in
deciding which variant is necessary for a desired accuracy. For the
Ewald formula, eqn. \ref{eq1.7}, convergence of the k-space part
clearly depends on the value of $G\Delta$, where $\Delta$ is the grid
spacing corresponding to the largest $k$ vector.

 This is also true for primitive
P$^{3}$M, where $1/G\Delta$ corresponds to the number of grid points
within
a Gaussian particle and thus controls the discretization error.
For P$^{3}$M where the Gaussian density is replaced
in the particle-mesh calculation
by an assignment function with a modified Coulomb Green's function
the situation is less clear intuitively . Figure 1 shows this
dependence on $G\Delta$.

   Figure 1 shows the relative accuracy of the k-space part of
the  forces for the various assignment schemes and for primitive
P$^{3}$M.(The results graphed are for 512 randomly placed charges in
a periodic cube and  a $32^{3}$ grid was used, but the
semi-quantitative features of the figure are insensitive to these
details.) The reference forces  used in determining the relative error,
for this figure were from a well converged Ewald calculation.

Several observations can be made from the results of figure 1:
\begin{itemize}
\item  Primitive P$^{3}$M (labeled ``S3'') is here the most accurate
but its
       drawback is the large number of grid points (shown above the
x-axis)
       required to describe the Gaussian if $G$ is small. (The  number
of
       grid points displayed corresponds to a cutoff of $10^{-7}$ on
the
       Gaussian density).  By contrast,
       for the various assignment schemes (Labeled by n)  only
       n$^{3}$ grid points per charge are needed. (Since the charge
density
       assignment factors as
       a product of x,y, and z values most of the operations involved
       in assigning the charge to the grid scale as n rather than
n$^{3}$.)

\item Each increase in $n$ gives almost an order of magnitude increase
in
      relative accuracy. Most condensed matter simulations aim for a
relative
      accuracy of at least $10^{-4}$. Although present day biomolecular
      force fields are very approximate this sort of accuracy would be
desirable
      for potential or free energy comparisons between configurations
or phases.

\item The results shown are for a random configuration of charges.
       Since the Hockney and Eastwood formula for the modified Coulomb
      Green's function
      is based on such a configuration it might be suspected that these
      configurations would give the highest accuracy. This is true but
even
      for a highly ordered configuration (perturbed lattice) the
reduction
      in accuracy was less than a factor of two.

\item  As mentioned above the figure is insensitive to the number of
       particles in the periodic cell and the grid size. For example,
       the same plot
       for a $64^{3}$ grid looks very similar.
\end{itemize}

\subsection{ Timing Comparisons}

       The Ewald, P$^{3}$M, and FMM algorithms have been used to
calculate
the forces and potential energy for a random periodic configuration of
N=512, 1000, 5000, 10000, and 20000 charges.
Total timings and other details are given in Table I.
The distribution of the total time over the various
operations is shown for the P$^{3}$M method in Table II and for FMM in
Table III.  The computations were done on an IBM RS/6000 590
workstation using the ESSL mathematical subroutine package for the FFT.

         Such comparisons depend strongly, of course, on the effort put
into optimizing the coding. As already discussed this is particularly
true of the FMM algorithm. These comparisons are therefore only
semi-quantitative. With this caveat in mind several trends are worth
 noting:
\begin{itemize}
\item The P$^{3}$M is roughly four times faster than the FMM algorithm
      for all N shown in the table. (An extrapolation based on the
      scalings for the two methods suggest that FMM only becomes faster
      at some unphysical size $N > 10^{60}$.)
\item Even for the N=512 system the P$^{3}$M is faster than
      the Ewald method.
      (We estimate this crossover occurs for $N\leq 50$).
      The FMM code used here is faster than the
      Ewald for $\approx 800$ particles. This indicates the
       optimization effort put into this code since crossovers
       in excess of 50,000 particles have been reported for
       these two methods.
\end{itemize}

\subsection{ Other Contrasts between P$^{3}$M and FMM}
        Although P$^{3}$M is faster than FMM  the choice between
        the two may be dictated by other considerations. These include:
\begin{itemize}
\item {\em Ease of Coding}\\
       The P$^{3}$M method is considerably easier to code than the FMM.
       The Poisson solver for periodic systems is based on available
       FFT software. The assignment of the particle charge to the grid
       and the interpolation of the electric field from the grid to the
       particles are both straight forward loops over the particles.
\item {\em Non-cubic periodic cells}\\
       Treating non-cubic, parallelepiped periodic cells is straight
       forward with the P$^{3}$M method using the corresponding
non-cubic
       grid.  This situation is more complicated for the FMM since
       the convergence of the expansions depends on the ratio of
distance
       from the origin to the distance of the nearest charge included
       in the expansion. For cells separated by one or more intervening
       cells this ratio is now anisotropic and, at the least,
additional
       bookkeeping is required.
\item {\em Alternate Boundary Conditions}\\
       The simulation of a cluster of charges (vacuum rather than
       periodic boundary conditions) or a slab periodic in only two
       dimensions is natural with the FMM and only
       involves omitting a step (discussed under step 3b in section 3).

       The P$^{3}$M algorithm can also be modified to treat these cases
by
       cutting off the Coulomb potential (either spherically for a
       cluster or in the transverse direction for the slab) at a
distance
       large enough to correctly include all interactions in the
       cluster or slab but short enough to eliminate interactions with
       any periodic images as suggested in \cite {hockney} and
       \cite{shimada}.

       Specifically the development of section two for a cluster
       or slab proceeds to eqn.~\ref{eq1.4} exactly as before and
       again the remaining integral is to be evaluated numerically
       by Fourier methods. As already stated this is done by first
       cutting off the Coulomb potential at $|r-r'|$ sufficiently large
       that $\hat{\rho}(r)\hat{\rho}(r')$ is zero and taking a periodic
       cell large enough to insure that the periodic images implied
       by a Fourier treatment do not interact. Note however that
       the density $\hat{\rho}(r)$ consists of Gaussians on
       each particle and thus extends somewhat beyond the cluster
       or slab boundaries.

       For a spherical cluster the potential can be cutoff at a
       distance $R_{c}$, which must exceed the cluster diameter by
several
       Gaussian widths, $1/\sqrt{2G}$, in order to include all
interactions
      within the cluster. The $4\pi/k^{2}$ in
      eqn.~\ref{eqA.1} is then multiplied by the form factor
      $[1-\cos(kR_{c})]$
       but otherwise unchanged. Taking the periodic cell length as
       $2R_{c}$ or greater, to avoid interactions between periodic
       images the computation proceeds as before with similar accuracy.

       For the slab the Coulomb potential can be cutoff when $z$
exceeds
        $Z_{c}$ which must exceed the slab width, as before, by
       several Gaussian widths. The $4\pi/k^{2}$ in eqn.~\ref{eqA.1} is
       now multiplied by the form factor
\beq
       1-\exp(-k_{\perp}Z_{c})\left(\cos(k_{z}Z_{c})-{k_{z}\over
k_{\perp}}
       \sin(k_{z}Z_{c})\right)
\eeq
      where $k_{\perp}=\sqrt{k_{x}^{2}+k_{y}^{2}}$. The last term,
      $(k_{z}/k_{\perp})\sin(k_{z}Z_{c})$, is
      singular at $k_{\perp}=0$ but this term can be shown not to
       contribute to the potential as $k_{\perp}\rightarrow 0$ and
      can be omitted for those $k$ values. (Of course, the $k=0$ value
      contributes nothing for charge neutral systems). With this
      procedure and again taking the periodic cell length along z
      large enough that the slab images are separated by at least
$Z_{c}$
      the method gives accuracy comparable to the fully periodic case.

      In sum, treating a cluster or slab with the $P^{3}$M algorithm
      involves a one time
      modification to the optimal Coulomb's Green's function and the
      use of a cell somewhat more than twice as large as the system
      dimension in either three or one dimensions.
      This only affects the FFT times and for similar accuracy the
      P$^{3}$M is still considerably faster than FMM.

\item {\em Non-Coulombic Interactions}\\
       The P$^{3}$M method proceeds similarly for any isotropic
       Fourier transformable
       potential. (Short-range repulsion can be treated separately so
       that the remainder is transformable).
       For a dipolar potential the vector Gaussian dipolar  density is
       assigned to the grid and the electric field again computed by
       Fourier methods.
\item {\em Dimensionality}\\
       Two dimensional systems may be easily treated by both
algorithms.
\end{itemize}
\section{Parallel Implementation of Ewald and P$^{3}$M}
        In this section parallel implementation of the k-space part of
eqns.~\ref{eq1.7} and \ref{eq1.4} is discussed for the Ewald and the
P$^{3}$M method. The real-space part is a reasonably short-ranged
pair interaction and parallel algorithms for these interactions have
been extensively discussed in the literature \cite{beazley}.
 The results we use for
illustration
were computed on the T3D using shared memory constructs
but the remarks  apply almost unchanged to any distributed memory
machine
or message passing system.
For discussion of a parallel FMM see~\cite{board}

\subsection{Ewald}
           This method is ideal for parallel implementation as very
little
interprocessor communication is required and several implementations
have been presented \cite{deLeeuw}. The required structure factors
are rewritten as
\beq
S(k)=\sum_{j} Z_{j}e^{ik\cdot r_{j}}=\sum_{P}\sum_{j\in P} e^{ik\cdot
r_{j}}
=\sum_{P} S_{P}(k)
                     \label{eq5.1}
\eeq
where $P$ denotes processors.
   Particles are distributed to processors and the partial structure
factor
$ S_{P}(k)$ computed on each processor for all $k$ vectors.
These  partial structure factors are then summed across processors
to obtain the total $S(k)$. The sum over $k$ vectors (eqn.~\ref{eq1.7})
to obtain the total energy  $U$ can be done on one or all processors.
 Computing the forces
(where the summand $|S(k)|^{2}$ is instead $i{\bf k}e^{ik\cdot r_{j}}
S(k)$)
involves no further interprocessor communication.
It is preferable to divide the particles among processors and
 have each processor handle all $k$ vectors since
then the usual complex multiplication can most easily be used to build
up the
necessary
$e^{i(k_{1}+k_{2})\cdot r_{j}}=e^{ik_{1}\cdot r_{j}}e^{ik_{2}\cdot
r_{j}}$.

 Figure 2 shows the time required to evaluate the energy and forces
versus the number of processors for
N=10240 particles and $G\Omega^{1/3}=8.3$.
Several observations can be
made:
\begin{itemize}
\item As already stated, scaling of the time with number of processors
      is almost ideal ($\propto 1/NPE$). The interprocessor
communication
      time was always significantly less than  $.1\%$ of the total.
\item If $G\Omega^{1/3}$ is scaled as $N^{1/6}$ to minimize the time
      (discussion in section 2) then
      the timings scale as $N^{3/2}$. Results for N=1024,
$G\Omega^{1/3}=5.6$,
      and N=102400, $G\Omega^{1/3}=12.2$ are not shown since the scaled
      timings overlap the results shown.
\end{itemize}

   For systems of the order of $10^{4}$ charges and a hundred
processors
the Ewald method is a good choice. For larger systems the
$N^{3/2}$ Ewald scaling dictates that the $P^{3}M$ algorithm should be
used.

\subsection{P$^{3}$M}

      Our replicated data, parallel version of the P$^{3}$M algorithm
follows
the steps listed at the end of section~2.

\begin{itemize}
\item    A discretized effective density
\beq
      W(r_{g})=\sum_{j=1}^{N}W_{j}(r_{g})=\sum_{P}\sum_{j\in P}
W_{j}(r_{g})
\eeq
 is formed at the grid points, denoted by $r_{g}$, by first summing the
contribution of particles on each processor and then summing over
processors to get the total effective density which we store on all
processors.

\item  This density is then partitioned for use with a distributed data
FFT. Since the effective density is stored on each processor no
interprocessor communication is required here.
 The total energy is calculated by first summing the distributed
Fourier components of the effective density and potential on each
processor
and then summing over processors. To compute the electrostatic force
on the particles (eqn. \ref{eq1.8}) the electric field on the grid
is reassembled (on the T3D a shared memory construct, shmem\_fcollect,
does this operation) and stored on each processor.

\item  Finally eqn.~\ref{eq1.8}, with $W$ replacing $\rho$, is
evaluated
for the particles on each processor to get the forces.
To save memory the same array was used to successively store the
electric field components although this requires duplicating
the single particle density calculations increasing the overall time
by roughly 20 \%.
\end{itemize}

    This replicated data implementation has several advantages: The
coding
is straightforward and all message passing is hidden in global sums or
other supplied routines (e.g. shmem\_fcollect); The assignment of
particles
to processors is arbitrary although a comparable number per processor
is
desirable for load balance. It has the disadvantage of requiring more
memory
than a purely distributed data implementation and
involves interprocessor communication (e.g. in forming the
total density and the electric fields) that could be minimized if the
particles were initially sorted on processors in a domain decomposition
corresponding to the grid decomposition used by the FFT. As seen in the
illustrative timings shown below this extra effort would be worthwhile
for smaller systems on a large number of processors where the
fraction of time used in interprocessor communications is significant.
The replicated data version is thus limited to less than a few hundred
processors.
The method is clearly well adapted to a distributed memory, domain
decomposition approach to remove this limit.
In this parallel version advantage has also not been taken in the FFTs
of the fact that the charge density and the electric fields are real.

    Figure 3 illustrates timing trends for the $n=4$ assignment
scheme for systems of 102,400 and 1,024,000 particles. These timings
are only
indicative with the same $64^{3}$ grid used for both systems.
The total time (upper solid line)  is broken into five, cumulative,
 components:
\begin{itemize}
\item  The time required to tabulate the effective density, $W(r_{g})$,
(lower
   solid line indicated by {\em Density} arrow) due to the particles on
   each
   processor. This step involves no communication and scales linearly
   with the number of particles per processor.
\item The time to get the total $W(r_{g})$ by summing across processors
      (second solid line)
\item The time spent in the FFT calculations for the discretized
potential
       and electric field terms (solid line indicated by {\em Poisson}
        arrow)
\item  The time spent summing Fourier components to get the
contribution
      to the total electrostatic energy. This time is not significant
      and is indiscernible in the figure  appearing as a broadening
      of the previous solid line.
\item Finally the time spent calculating the forces (interpolating the
      electric field from the grid to the particle positions) is
      indicated on the right by the $E_{i}$ arrow. This time
      is roughly three times that required to form $W(r_{g})$ since
      similar steps are required but now for three components.
      Again no interprocessor communication is involved and this step
      scales ideally.
\end{itemize}
   The dashed line shows the interprocessor communication time from
   component two (summing $W(r_{g})$ across processors) and collecting
   the electric field components included in component five.
   (Communication time from the FFTs is included in {\em Poisson}).
    As the time for the
   rest of the calculation decreases linearly with the number of
   processors
   this slowly increasing component, (it almost doubles in going from 8
to
   256 processors)  becomes important rising to 75\%
   of the total time for N=102,400 and 256 processors. For  N=1,024,000
   the communications times are about the same but less important.

   For 102,400 charges and 32 processors the P$^{3}$M method is
approximately 50 times
faster than Ewald and is several hundred times faster for the
1,024,000 charge case.

\section{Summary}

       In this paper it has been argued that the Ewald method is
suitable
for systems of a few hundred particles per processor.
For larger systems the  $P^{3}M$ algorithm  is increasingly more
efficient.
The FMM is a
second choice for all system sizes both in terms of speed and program
complexity. Considerations such as the relative advantages gained in
not
updating the field due to distant particles every molecular dynamics
step
or the use of accelerated series convergence methods \cite{lustig} with
the
FMM have not been addressed but seem unlikely to alter these
conclusions.

\acknowledgements

   We thank Steve Plimpton of Sandia National Laboratory, Terry Stouch
and Malcolm Davis
of Bristol-Myers Squibb, Brock Luty of ETH, and Jim Belak of LLNL for
their
comments and valuable suggestions. We particularly thank Kouchik Ghosh
of
Cray Research for the use of his distributed data FFT code on the T3D.
Work done at Lawrence Livermore National Laboratory is supported by the
U. S. Department of Energy under Contract W-4705-Eng-48.
\appendix
\section{P$^{3}$M Formulae}
      Some useful formulas for the P$^3$M method  are collected here
starting
with the assignment functions, $W_{n}(x)$, used to create a density and
to
interpolate the forces. The subscript on $W_{n}$(x) refers to
the number of grid points (along each coordinate axis)
 ``supporting'' a  charged particle
and $x$ is the distance from the particle to the grid point measured
in grid spacings. The full assignment function is the product
along each direction, $W(r)=W(x)W(y)W(z)$. The $W_{n}$ are zero outside
the
indicated range and are given by successive convolutions,
 $W_{n+1}=W_{n}\ast W_{1}$.
These assignment functions satisfy
\beq
\begin{array}{ccc}
\int W_{n}(x)dx=1&  \mbox{and}  & \sum_{j=-\infty}^{\infty}W_{n}(x+j)=1
\end{array}
 \eeq
The first five $W_{n}(x)$ are:

\beq
\begin{array}{l}
\begin{array}{ll}W_{1}(x)=1&  \mbox{$|x|<1/2$} \end{array}
\\\\

\begin{array}{ll}W_{2}(x)=1-|x|&  \mbox{$|x|<1$} \end{array}
\\\\

W_{3}(x)=\left\{   \begin{array}{ll}
                  3/4-x^{2}&     \mbox{$|x|<1/2$}\\
                  {1\over 2}(3/2-|x|)^{2}&     \mbox{$1/2<|x|<3/2$}
                  \end{array}
         \right.
\\\\

W_{4}(x)=\left\{   \begin{array}{ll}
                  2/3-x^{2}+|x|^{3}/2&     \mbox{$|x|<1$}\\
                  (2-|x|)^{3}/6&     \mbox{$1<|x|<2$}
                  \end{array}
         \right.
\\\\

W_{5}(x)=\left\{   \begin{array}{ll}
                (115-120 x^{2}+48 x^{4})/192  &     \mbox{$|x|<1/2$}\\
                (55+20|x|-120 x^{2}+80|x|^{3}-16 x^{4})/96  &
\mbox{$1/2<|x|<3/2$}\\
                (-5+2|x|)^{4}/384                & \mbox{$3/2<|x|<5/2$}
                 \end{array}
         \right.
\end{array}
\eeq

      Higher order functions, if needed, can be easily obtained as
\beq
W_{n}(x)=\sum_{l=0}^{n-1} A_{n}(l,j)(x-j/2)^{l}\;\;\;\mbox{for}\;\;\;
       -{1\over 2}<x-{j\over 2}<{1\over 2}
\eeq
where $j$ goes from $-(n-1)$ to $(n+1)$ in steps of 2. The $A_{n}(l,j)$
satisfy the recursion relations:
\beq
\begin{array}{l}
A_{n+1}(l+1,j)=\begin{array}{c}\underline{A_{n}(l,j+1)-A_{n}(l,j-1)}\\
l+1
                \end{array}\\
\displaystyle
A_{n+1}(0,j)=\sum_{l=0}^{n-1} ( 2)^{-l}\begin{array}{c}\underline{
     [A_{n}(l,j-1)+(-1)^{l}A_{n}(l,j+1)]}\\ l+1\end{array}
\end{array}
\eeq
starting from $A_{1}(0,0)=1.0\;$.

To minimize the error associated with using the assignment function
$W(r)$
in lieu of $\hat{\rho}(r)$ as well as aliasing errors Hockney and
Eastwood
derive for the modified Coulomb Green's function \cite{ferrell}
\beq
G(k)=\begin{array}{c}4\pi\\\overline{k^{2}}\end{array}
\begin{array}{c}\underline{\displaystyle \sum_{\bf b}
       \begin{array}{c}\underline{ {\bf k}\cdot(\bf{k+b})}\\
                                   |\bf{k+b}|^{2} \end{array}
     W_{n}^{2}(\bf{k+b})\hat{\rho}^{2}(\bf{k+b})}\\
\displaystyle
      \left[ \sum_{\bf b} W_{n}^{2}(\bf{k+b}) \right]^{2}
\end{array}
                          \label{eqA.1}
\eeq
where the usual Fourier wave vectors,
${\bf k}=2\pi {\bf j}/L$ with $j=1\ldots M$ for $M$ grid points on a
periodic cell length $L$ and the Brillouin zone vectors ${\bf
b}=2\pi{\bf l}/
\Delta $ where $l=-\infty,\infty$ and $\Delta=L/M$.
This expression needs to be evaluated only once.

   Some ingredients in this expression are the Fourier transform
for the assignment function
\beq
W_{n}(k)=W_{1}(k)^{n}=\left(
\begin{array}{c}\underline{\sin(k\Delta/2)}\\k\Delta/2\end{array}
                       \right)^{n}
\eeq
and
\beq
\hat{\rho}(k)=e^{-k^{2}/4G^{2}}\;.
\eeq

The sum in the denominator
is evaluated using the identity
\beq
\cot(x)=\sum_{j=-\infty}^{\infty}{1\over  x+\pi j}
\eeq
introduced by Hockney and Eastwood, so
\beq
\begin{array}{ll}
S_{n}(k)&\equiv   \displaystyle
\sum_{b}W_{n}^{2}(k+b)=[\sin(k\Delta/2)]^{2n}
\displaystyle
   \sum_{j=-\infty}^{\infty}\begin{array}{c} 1\\ \overline{\left[
k\Delta/2+\pi j\right]^{2n}}\end{array}\\
&\\
&=
[\sin(k\Delta/2)]^{2n}\left[
\begin{array}{c} -1\\\overline{(2n-1)!}\end{array}\left(
\begin{array}{c}d^{2n-1}\\\overline{dx^{2n-1}}\end{array}
 \cot(x)\right|_{x=k\Delta/2}\right]\;.
\end{array}
\eeq
Denoting $z=\sin(k\Delta/2)$ some algebra gives
\beq
S_{n}(k)=\left\{
\begin{array}{ll}
(1-2z^{2}/3)\;\;\;\;\;\;&\mbox{$n=2$}\\
(1-z^{2}+2 z^{4}/15) &\mbox{$n=3$}\\
(1-4 z^{2}/3+2 z^{4}/5+4 z^{6}/315)  & \mbox{$n=4$}\\
(1-5 z^{2}/3+7 z^{4}/9-17 z^{6}/189 +2 z^{8}/2835) & \mbox{$n=5$}
\end{array}
\right.  \;.
\eeq
Higher order $S_{n}(k)$, if needed, are given by
\beq
S_{n}(k)=\sum_{l=0}^{n-1}b_{n}(l) z^{2l}/\Gamma(2n)
\eeq
where the $b_{n}(l)$ satisfy:
\beq
\begin{array}{l}
b_{n}(l)=4[b_{n-1}(l)(n-l-1)(n-l-1/2)-b_{n-1}(l)(n-l)^{2}]    \\
b_{n}(0)=4b_{n-1}(0)(n-1)(n-1/2)
\end{array}
\eeq
starting from $b_{1}(0)=1$.

The full $S_{n}({\bf k})=S_{n}(k_{x})S_{n}(k_{y})S_{n}(k_{z})$.
The numerator of eqn. \ref{eqA.1} converges rapidly and is
evaluated numerically.
\section{Timing and Selection of Parameters for P$^{3}$M }

      The time required to compute the total energy and the electric
field at
each particle in an N particle system using P$^{3}$M with a real space
cutoff of $r_{c}$, order n assignment scheme, and a grid of $N_{g}$
points may be expressed as
\beq
T=a_{1}\left({4\pi\over 3}\rho r_{c}^{3}\right)
N+\left(a_{2}N+a_{3}Nn^{3}
       \right) +(a_{4}N_{g}\ln N_{g}+a_{5}N_{g})\;.
                  \label{eqa2.0}
\eeq
The first term gives the time for the particle-particle interactions,
the
second term the time to form the density and to interpolate the fields
from the grid to the particle positions, and the last term the time
to do the FFTs.

     Assuming that a desired precision is specified, the allowed
real space (particle-particle) error, $\epsilon_{R}(Gr_{c})$,
(which to a good approximation can be fitted by $\epsilon_{R}\sim
\exp(-G^{2}
r_{c}^{2}/2)/\sqrt{Gr_{c}}\;$~),
gives $Gr_{c}=c_{R}(\epsilon_{R})$  and the allowed k-space error,
$\epsilon_{k}=F[G\Delta,n]$, (see figure 1), implies
$G\Delta=c_{k}(n,\epsilon_{k})$ or $G^{3}\Omega/N_{g}=c^{3}_{k}$.
Using these relations to express $r_{c}$ and
$N_{g}$ in terms of $G$, the time can now  be varied to find the
optimal $G$
and $n$.

Variation with $G$, $\partial T/\partial G=0$, implies
\beq
a_{1}{4\pi\over 3}\rho{c_{R}^{3}\over G^{6}}N=a_{4}{\Omega\over
c_{k}^{3}}
\left[\ln{G^{3}\Omega\over c_{k}^{3}}+1\right]+a_{5} {\Omega\over
c_{k}^{3}}
                  \label{eqa2.1}
\eeq
so
\beq
G=\left[\begin{array}{c}\underline{a_{1}{4\pi\over 3}\rho^{2} c_{R}^{3}
       c_{k}^{3}(n)}\\
       a_{5}+a_{4}[\ln{G^{3}\Omega\over
c_{k}^{3}}+1]\end{array}\right]^{1/6}
               \label{eqa2.2}
\eeq
which, for a given $n$, can be quickly iterated to find $G$.
The optimal $n$ can be obtained by direct evaluation of $T$ at the
optimal
$G$ for $n=3,4,5\ldots$.

   If the logarithmic variation of the FFT time is ignored,
eqn.~\ref{eqa2.1}
implies $T_{R}\approx T_{FFT}$ at the optimal $G$. This gives a rule of
thumb
for adjusting $G$
\beq
G_{new}\approx G_{old}(T_{R}^{old}/T_{FFT}^{old})^{1/6}
\eeq
with corresponding adjustments in $r_{c}$ and $N_{g}$ to maintain
accuracy.
Equations \ref{eqa2.0} and \ref{eqa2.2} also imply that
$T\sim N [a+b\ln N]^{1/2}$ \cite{hockney2}.

  For many systems there are also moderate range forces
 (e.g. Van der Waals) to be calculated and the $r_{c}$ may be specified
by
these forces. The allowable real space error then determines $G$ and
the k-space error determines the necessary grid size  $N_{g}$. The
optimal
$n$ is determined empirically unless the $\{a_{1}\ldots a_{5}\}$ are
known.
\section{Effective implementation of FMM}
      Efficient implementation of the FMM typically requires more
effort
than for the more straight forward Ewald or P$^{3}$M methods. We
discuss in this appendix some of the technical details.

   The expansion order $p\equiv l_{max}+1$ is determined by the
required
accuracy $\epsilon$ and the number of subdivision levels $L$ is then
varied to minimize the computing time.
Truncation of the various expansions gives an error estimate of the
form $\epsilon\approx c\alpha^{p}$, where the coefficient c and
the geometry related ratio $\alpha$ have been determined empirically,
to give the following relation for  the required expansion order
necessary for a desired accuracy
\beq
    p \approx -\ln(100\epsilon )/\ln(2) \;.
\eeq

The total time is the sum of the times for the five steps listed in the
section III.
\begin{itemize}
\item
    In step 1 the time to evaluate all multipoles at the finest
subdivision
\beq
   T_{1} = a_{1} N p (p+1)/2
\eeq
where $a_{1}$ is the time to evaluate one term in Eq. \ref{eq3.2}.
(Note the $Y_{lm}$ s can be calculated recursively.)

\item
In Step 2, Eq. 18 is use to shift the multipoles of the child
(finer subdivision) to the box
center of the parent (coarser subdivision). This is done for each box
at all
levels except for the
highest level box.
\beq
T_{2} = {a_{2}\over 4} p^{2}  (p+1)^{2}   (N_{boxes}-1)
\eeq
where $N_{boxes}$ is the total number of boxes in the hierarchical tree
(all levels)
($N_{boxes}=( 8^{L+1}-1)/7\;\;$), and $a_{2}$ equals the time for one
complex
multiply and add.

\item
The time in Step 3a using Eq. \ref{eq3.10} to transfer the local
expansion of
 the parent to the center of the child
\beq
T_{3a} =  {a_{2}\over 4} p^{2}  (p+1)^{2}  (N_{boxes}-1)
\eeq

\item
 The time in step 3b to convert the multipole expansion of the members
of the
interaction list (at most 189 members) of to local expansion about the
center
of that box using Eq. \ref{eq3.12}
\beq
T_{3b} = {a_{2}\over 2} p^{3}  (p+1)  189  (N_{boxes}-1)
\eeq

\item
The time required in step 4 to evaluate the local expansion (Eq.
\ref{eq3.3}
for each particle in the system
\beq
T_{4} = {a_{1}\over 2}N p(p+1)
\eeq

\item
Finally in step 5, the near field interaction is evaluated by summing
up over
all pairs in neighbors boxes.
\beq
T_{5} = a_{3}  {27\over 2}  N {N\over 8^{L}}
\eeq
where $a_{3}$ = time to evaluate one pair interaction.
\end{itemize}

The total time is thus,
\beq
T= a_{1}Np(p+1) + [{a_{2}\over
2}p^{2}(p+1)(p+1+189p)](N_{boxes}-1)+a_{3}
 {27\over 2} N {N\over 8^{L}}
\eeq

Minimizing with $L$ gives an expression for the optimal value
\beq
8^{L_{opt}}   = N\sqrt{
\begin{array}{l} 27(a_{3}/a_{2})(7/8) \\
       \overline{p^{2}(p+1)(p+1+189p)} \end{array}
        }\;.
\eeq

This gives an optimal time of
\beq
T_{opt} = N\left[a_{1}p(p+1) +
\sqrt{(216/7)a_{2}a_{3}p^{2}(p+1)(190p+1)} \right]
\eeq
(where we have approximated $N_{boxes}-1\approx 8^{L+1}/7$).

Both terms in $T_{opt}$ scale as $p^{2}$ however the coefficient for
the
second term is
much larger. In fact at $L=L_{opt}$ almost all the time is divided
equally
between converting multiples to local expansions (step 3b) and the
direct
coulomb sums (step 5).

Greengard observed that the transformation equations
\ref{eq3.8}, \ref{eq3.10}, \ref{eq3.12} are all of the form of a
convolution  and  FFT's could be used to speed up their evaluation.
To use Fourier transforms to evaluated these equations
$ t^{MM}$, $t^{LL}$, and $t^{LM}$
must be mapped to periodic function in such a way as not to change
the the rhs of these equations.  This can be done by padding these
vectors
with zeros.  For example define $M_{p}(l,m)$ for l in [0,2p-1] and m
in [-2p,2p-1] as,
\beq
      M_{p}(l,m)   =\left\{ \begin{array}{l}  M(l,m)\;\;\;\mbox{for }
                                                      \;\;\;
                                                      |m|<=l\;\;, l<p\\
                                            0\;\;  \mbox{otherwise}
                         \end{array}  \right.
\eeq

Using FFT's equations \ref{eq3.8}, \ref{eq3.10}, \ref{eq3.12}
 in Fourier space will take

\[  T_{2}  =T_{3a}= a_{2} (2p)^{2}  (N_{boxes}-1)     \]
\[ T_{3b} = a_{2} (2p)^{2} 189(N_{boxes}-1)   \]

(Note we used the symmetry $M(l,-m) = (-1)^{m}  M^{*}(l,m)\;$) which
in Fourier space
implies $M_{p}^{k}(l+2p,-m) = M_{p}^{k}(p,m)\;$), This reduces
the number of terms to evaluate by a factor of two)

Steps 2, 3a and 3b can all be formed in Fourier space, so in principle
all
that is needed is to transform to multipole coefficient M at the finest
level
to Fourier space and the local expansion coefficient L also at the
finest
level back to real space.   The time to do this, using symmetry is,
\beq
T_{fft} =  a_{2}  2  [8p^{2}  \ln(2p) ]  8^{L}
\eeq

Before  writing down the timing for the FMM algorithm using FFT's a
numerical
issue needs to be considered.  The vectors ${\cal M}$, ${\cal L}$,
$ t^{MM}$,$t^{LL}$, and $t^{LM}$ are factorially
varying functions of their indices.  This large dynamic range results
in lost
of precision when performing the FFT and hence the Fourier formulation
of
equations \ref{eq3.8}, \ref{eq3.10}, \ref{eq3.12} becomes unstable for
large p.
 One way to reduce the
dynamic range is by scaling. For example consider equation
\ref{eq3.12}.
 If we introduce
a scaling factor s we can rewrite equation \ref{eq3.12}  as,
\beq
    {\cal L}^{'}_{l'm'} = \sum_{l,m} s^{(-l')}t^{LM}(l+l',m'-m)s^{l+l'}
                            s^{-l}{\cal M}_{lm}
\eeq

by introducing $L_{s}(l,m) = s^{l} {\cal L}^{'}_{l m} $,
 $M_{s}(l,m) = s^{-l}{\cal M}_{lm}$ and
$T_{s}(l,m) = s^{l}t^{LM}(l,m)$
\beq
    L_{s}(l',m') = \sum_{l,m} T_{s}(l+l',m'-m)  M_{s}(l,m)
\eeq

Now s can be chosen to minimize the dynamic range. The above
transformation
can also be performed in Fourier space, and will be more stable  than
the
original FFT formulation of equation 22. This however, is not a
complete cure
but will stabilize the algorithm up to p=16.  The same procedure can be
applied to equations \ref{eq3.8}  and \ref{eq3.10}, however a different
 scaling s must be chosen.
This will require Fourier transformation to be performed on each of the
scaled
functions.  These additional FFTs offset the benefit of performing
equations
\ref{eq3.8} and \ref{eq3.10} in Fourier space for small p.
 The cross over is at p=16 which is the
limit of stability of the FFT approach. In light of this these two
equations
are performed in real space.  It is however worth evaluating
\ref{eq3.12} in
Fourier space since the each FFT can be amortized over 189
transformations. The cross over for
chosing direct or Fourier space for step 3b is about p=2. For this
formulation
of the FMM algorithm Lopt is given by

\beq
8^{L_{opt}}   = N\sqrt{
\begin{array}{l} 27(a_{3}/a_{2})(7/8) \\
       \overline{p^{2}[(p+1)^{2}+8\cdot189]} \end{array}
        }
\eeq

with a $T_{opt}(N,p) $
\beq
T_{opt} = N\left[a_{1}p(p+1) +
\sqrt{(216/7)a_{2}a_{3}p^{2}[(p+1)^{2}+1512]} \right]
\eeq

Comparing the two optimal timing expressions the FFT approach should be
favorable for $p>2$.

\begin{table}[h]
\begin{tabular}{|c|c|c|c|c|c|c|}\hline
N & Ew \{G,k$_{max}\}$ &  P$^{3}$M \{n,G,N$_k$\} &FMM
\{levels,l$_{max}$
\}& time (sec)&
 F$_{rel. err.}$& U$_{rel. err.}$\\\hline
512  & & \{5,2.12,24\}&    &  .06     &   3x 10$^{-5}$  &  4x 10$^{-5}$
\\
512  & && \{1,7\}          &  .31     &   2x 10$^{-5}$  &  4x 10$^{-5}$
\\
512  & \{7.74, 8\}&  &     &  .25     &   3x 10$^{-5}$  &  1x 10$^{-5}$
\\\hline
1000 && \{5,2.11,30\}&     &  .10     &   4x 10$^{-5}$  &  8x 10$^{-5}$
\\
1000 & && \{2,7\}          &  .54     &   5x 10$^{-5}$  &  2x 10$^{-5}$
\\
1000 &  \{10.75, 8\}&&     &  .68     &   4x 10$^{-5}$  &  4x 10$^{-5}$
\\\hline
5000 && \{5,2.09,48\} &    &  .48     &   1x 10$^{-5}$  &  3x 10$^{-4}$
\\
5000 & && \{2,7\}          &  3.45    &   8x 10$^{-5}$  &  3x 10$^{-4}$
\\
5000 &  \{10.32, 11\}&&    &  8.02    &   8x 10$^{-5}$  &  3x 10$^{-5}$
\\\hline
10000&& \{5,2.09,64\} &    &  1.12    &   7x 10$^{-5}$  &  4x 10$^{-4}$
\\
10000&  && \{3,7\}         &  5.21    &   9x 10$^{-5}$  &  7x 10$^{-4}$
\\
10000 &  \{10.32, 11\}&&   &  26.0    &   6x 10$^{-5}$  &  9x 10$^{-5}$
\\\hline
20000&& \{5,2.08,80\} &    &  2.40    &   6x 10$^{-5}$  &  2x 10$^{-5}$
\\
20000&  && \{3,7\}         &  10.01   &   7x 10$^{-5}$  &  1x 10$^{-4}$
\\
20000 &  \{12.90, 14\}&&   &  78.0    &   5x 10$^{-5}$  &  3x 10$^{-5}$
\\\hline
\end{tabular}
\caption[tabone]
    { Timings for Ewald, P$^{3}$M, and FMM on periodic systems of
      various size. Parameters used in these runs are given  in
brackets
      $\{\ldots\}$. In P$^{3}$M \{n,G,N$_k$\} $n$ refers to the
assignment
      function used and $N_{k}^{3}$ is the grid size. In FMM
     \{levels,l$_{max}$\}
      the periodic cell is divided into $8^{levels}$ cells.}
\end{table}

\begin{table}[h]
\begin{tabular}{|c|c|c|c|c|c|c|}\hline
N   & \{n,G,N$_k$\} &  R space  & Make $\rho$ & Poisson & E$_i$&
Total\\ \hline
 512 & \{5,2.12,24\}   &  .02      &  .00        &  .02   & .02  &
.06\\ \hline
1000&  \{5,2.11,30\}   &  .04      &  .01        &  .03   & .02  & .10
\\\hline
5000& \{5,2.09,48\}    &  .17      &  .06        &  .14   & .11  & .48
\\\hline
10000& \{5,2.09,64\}   &  .36      &  .12        &  .38   & .26  & 1.12
\\\hline
20000& \{5,2.08,80\}   &  .73      &  .25        &  .95   & .47  & 2.40
\\\hline
\end{tabular}
\caption[tabtwo]
   {Breakdown of the total time for the P$^{3}$M computations of
table~I
    into time spent summing the short-range interactions (excluding
    time spent to construct near neighbor tables), time spent assigning
the
    charge to the grid, Solving the Poisson eqn. for the potential and
field
     on the grid, and the time spent computing the total potential and
    interpolating the field to the particle locations.}
\end{table}

\begin{table}[h]
\begin{tabular}{|c|c|c|c|c|c|c|c|}\hline
 N &\{levels,l$_{max}$\} & tables & upward pass & downward pass & local
& near
 field & Total \\\hline
  512& \{1,7\}  & 0.00  & 0.02 & 0.04 & 0.01 & 0.24 & 0.31 \\\hline
 1000& \{2,7\}  & 0.02  & 0.07 & 0.30 & 0.02 & 0.13 & 0.54 \\\hline
 5000& \{2,7\}  & 0.03  & 0.14 & 0.31 & 0.12 & 2.82 & 3.45 \\\hline
10000& \{3,7\}  & 0.40  & 0.57 & 2.43 & 0.24 & 1.53 & 5.21 \\\hline
10000& \{3,7\}  & 0.42  & 0.77 & 2.45 & 0.48 & 5.82 & 10.01\\\hline
\end{tabular}
\caption[tabthree]
  { Breakdown of FMM times into the steps described in section 3. The
    initial category, tables, refers to time spent tabulating parentage
     for the cell hierarchy.}
\end{table}

\begin{center}
Figure Captions
\end{center}

\noindent
1.   The relative error in the forces,
   $\sqrt{\sum_{i=1}^{N}({\bf f}^{k}_{i}-{\bf f}_{i}^{k\;\;exact})^{2}/
   \sum_{i=1}^{N}({\bf f}_{i}^{total})^{2}}$ for 512 particles randomly
    located in the periodic cell for various assignment schemes plotted
    vs 1./``discretization'' (where ``discretization'', $1/G\Delta$,
    is proportional to the number of grid points under each Gaussian
    charge). The numbers above the x-axis indicate the number of
    grid points each charge is assigned to in the primitive P$^{3}$M
     method (curve labeled S3)\\

\noindent
2.   Timings for the k-space part of the Ewald sum performed on NPE=1
     up to 128 CRAY T3D processors for 10240 particles.\\

\noindent
3.   Timings for the PM step of the P$^{3}$M algorithm versus 1/ number
    of CRAY T3D processors for 102,400 and 1,024,000 particles.
    The time spent in assigning the density to
    the grid, solving the Poisson equation, and interpolating the
fields
    to the particles are indicated by labels at right. Gaps not
    labeled involve interprocessor communication times for density and
    field transfers whose sum is shown by the dashed line.
    A $64^{3}$ grid and n=4 assignment scheme was used for both
cases.\\

\begin{thebibliography}{99}
\bibitem[7]
{ewcomment}   Since the sum is only conditionally convergent these
            operations are uncontrolled. As this is only a motivational
            derivation and the results well known we ignore this. The
            usual formula derived here corresponds to no surface
            depolarization charge or ``tinfoil'' boundary conditions.
\end{thebibliography}
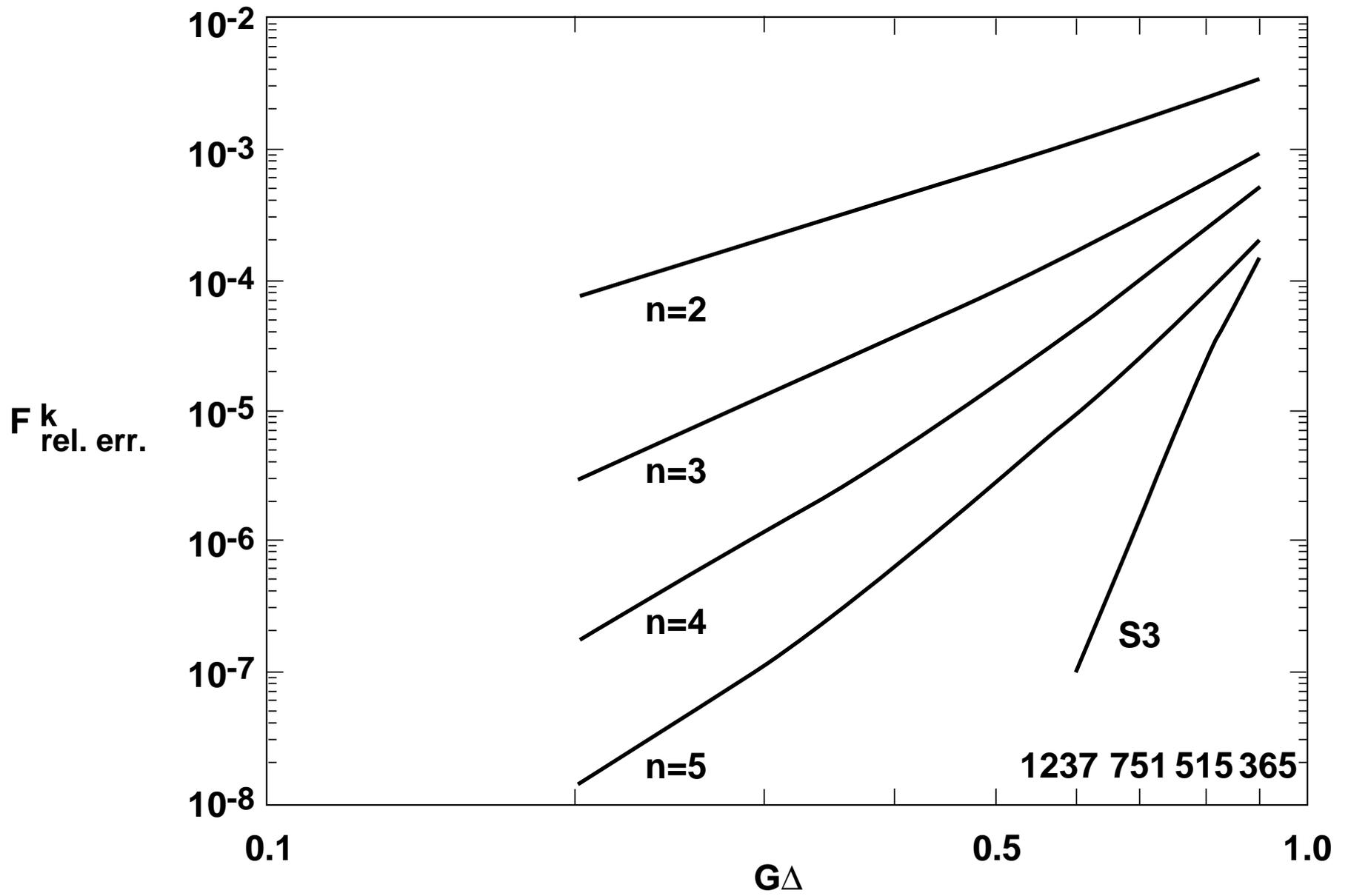
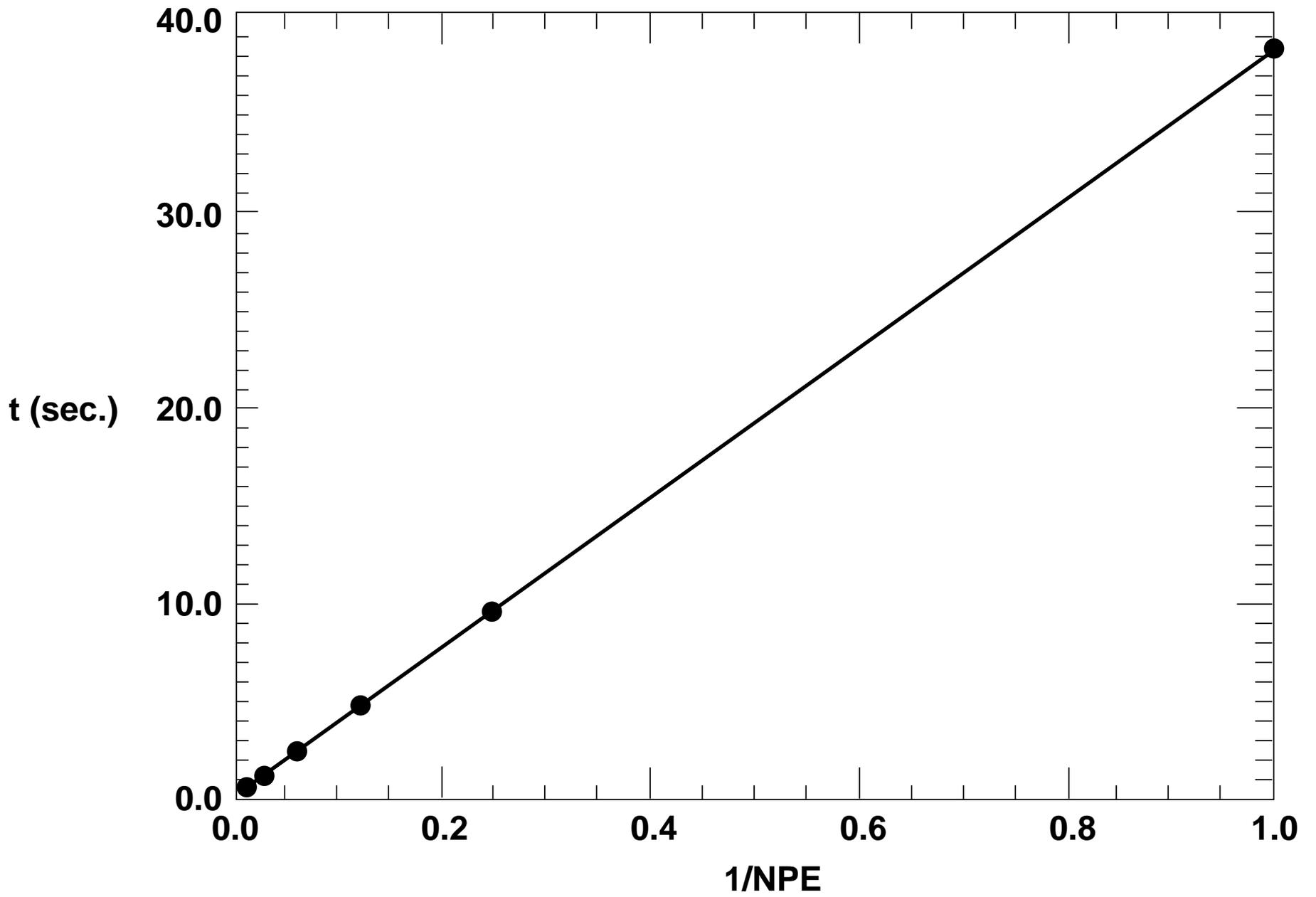
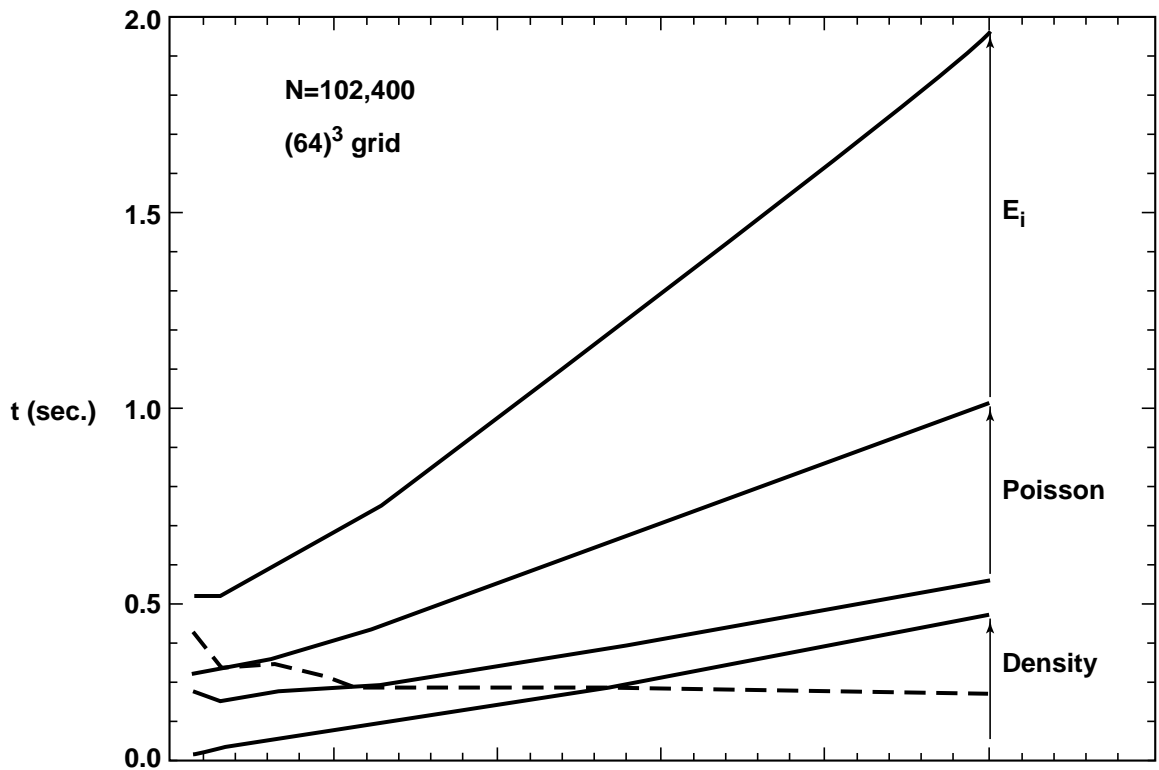
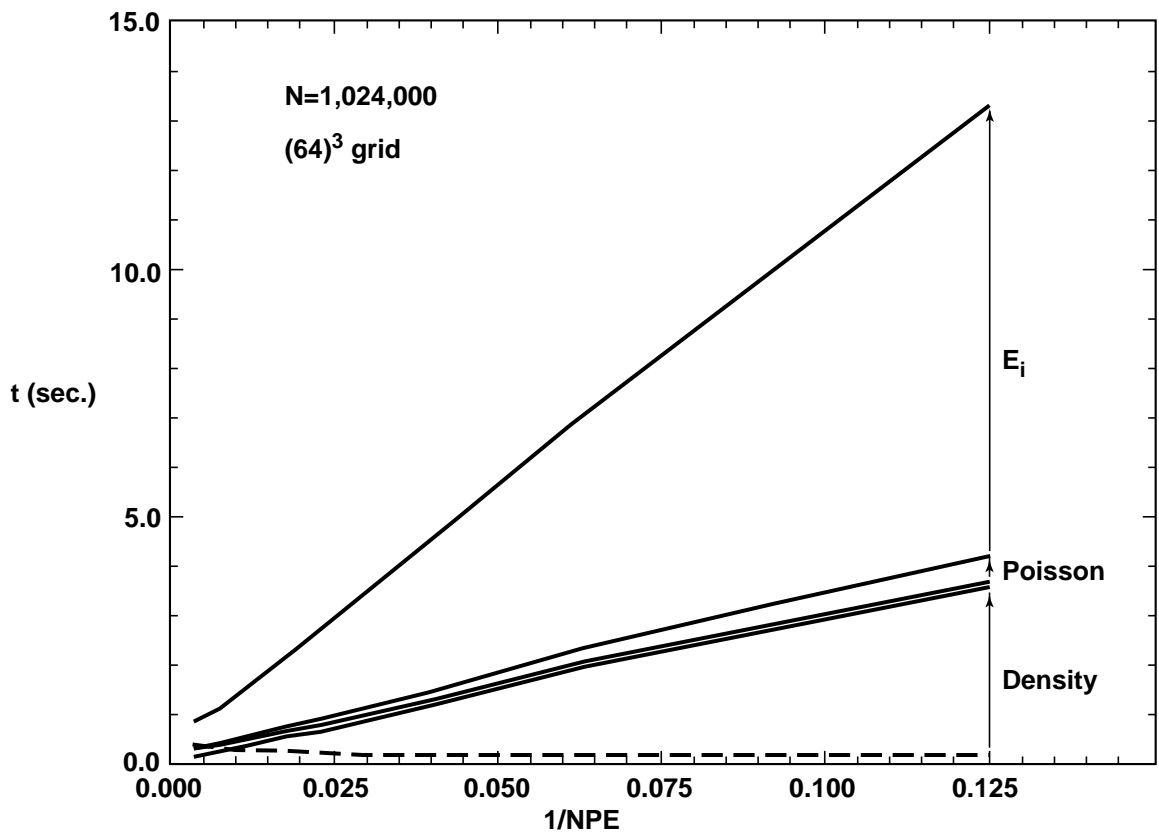
\end{document}